\documentclass[%
 reprint,
 superscriptaddress,
 amsmath,amssymb,
 aps,
 floatfix,
]{revtex4-2}

\usepackage{graphicx}
\usepackage{dcolumn}
\usepackage{bm}

\usepackage[utf8]{inputenc}
\usepackage[T1]{fontenc}
\usepackage{mathptmx}
\usepackage{etoolbox}
\usepackage{CJKutf8}
\usepackage[version=3]{mhchem}
\usepackage[separate-uncertainty = true,multi-part-units=single]{siunitx}
\usepackage{listings}
\usepackage{csquotes}
\usepackage{gensymb}
\usepackage[dvipsnames]{xcolor}

\usepackage[colorlinks=true,
            linkcolor=blue,
            citecolor=blue,
            urlcolor=blue]{hyperref}

\begin{document}

\title{Quantum computing for accurate large-scale electronic-structure calculations: DFT-embedded, post-processed quantum-selected configuration interaction}

\author{Tuan Minh Do}
\email{do.tuan.minh.qiqb@osaka-u.ac.jp}
\affiliation{Center for Quantum Information and Quantum Biology, The University of Osaka, 1-2 Machikaneyama, Toyonaka, Osaka 560-8531, Japan.}
\author{Yuichiro Yoshida}
\email{yoshida.yuichiro.qiqb@osaka-u.ac.jp}
\affiliation{Center for Quantum Information and Quantum Biology, The University of Osaka, 1-2 Machikaneyama, Toyonaka, Osaka 560-8531, Japan.}
\author{Tomoya Shiota}%
\affiliation{Center for Quantum Information and Quantum Biology, The University of Osaka, 1-2 Machikaneyama, Toyonaka, Osaka 560-8531, Japan.}
\author{Wataru Mizukami}
\email{mizukami.wataru.qiqb@osaka-u.ac.jp}
\affiliation{Center for Quantum Information and Quantum Biology, The University of Osaka, 1-2 Machikaneyama, Toyonaka, Osaka 560-8531, Japan.}
\affiliation{Graduate School of Engineering Science, The University of Osaka, 1-3 Machikaneyama, Toyonaka, Osaka 560-8531, Japan.}

\date{\today}

\begin{abstract}
We present a multilevel embedding framework for quantum chemistry calculations on a quantum computer. In our framework, a quantum algorithm treats the strongly correlated active space, while a high-level wave-function method such as coupled cluster theory or multireference perturbation theory recovers the remaining correlation in the surrounding region. A sampling-based quantum algorithm, quantum-selected configuration interaction, bridges the quantum and classical treatments. The entire calculation is embedded in a low-cost density functional theory description of the surrounding environment using Manby's projection technique. 
We apply the framework to organic, metal-organic, and metallic systems, computing bond dissociation energies, adsorption energies, and reaction barriers using only the subset of qubits of a 144-qubit superconducting quantum computer at the University of Osaka and achieving $\sim$\SI{1}{kcal/mol} agreement with classical references for a Menshutkin $\mathrm{S_N2}$ reaction inside a carbon nanotube.
Our results may open the way to quantitatively reliable quantum–classical hybrid calculations for large-scale chemical systems.
\end{abstract} 

\maketitle

\section{Introduction} \label{Introduction}
Over the past decade, quantum computing for quantum chemistry has begun to transition from a largely conceptual idea toward a practical tool for electronic-structure calculations.
On the hardware side, this progress is reflected in the increasing availability of early devices with 10$^2$--10$^3$ physical qubits, together with substantial improvements in device fidelity and operational reliability \cite{wurtzAquilaQuEras256qubit2023,abughanemIBMQuantumComputers2025,ransfordHelios98qubitTrappedion2025,winterspergerNeutralAtomQuantum2023,nguyenQuantumCloudComputing2024}.
On the software side, quantum algorithms have been integrated with classical electronic-structure methods to form hybrid quantum--classical algorithms. 
In these methods, a quantum computer targets a relatively small yet chemically critical subsystem, while the remaining part of the system is handled on classical hardware. Such hybrid approaches have moved beyond proof-of-concept demonstrations and have shown promising accuracy in increasingly realistic chemical settings \cite{khinevichEnhancingQuantumComputations2025,erhartCoupledClusterMethod2024,erhartCoupledClusterMethod2025,yoshidaAuxiliaryfieldQuantumMonte2025}.

Among current quantum--classical hybrid strategies, quantum-selected configuration interaction (QSCI) is one of the most practical approaches~\cite{kannoQuantumSelectedConfigurationInteraction2023}. In QSCI, a quantum algorithm such as quantum phase estimation or the variational quantum eigensolver (VQE) ~\cite{peruzzoVariationalEigenvalueSolver2014,tillyVariationalQuantumEigensolver2022} first prepares a targeted quantum state on a quantum computer. The prepared state is then sampled in the computational basis to identify the dominant Slater determinants of the molecular electronic wave function. These determinants define a compact subspace for constructing an effective Hamiltonian and performing subsequent simulations on classical hardware.
The efficient use of the measurement shots and the inherent noise resilience make QSCI a scalable and promising approach for practical quantum-chemical calculations. For example, Robledo-Moreno et al. used up to 77 physical qubits to perform QSCI calculations of $\ce{N2}$, $\ce{[Fe2-S2]}$, and $\ce{[Fe4-S4]}$ clusters~\cite{robledo-morenoChemistryScaleExact2025}. 
The use of QSCI and its variants is becoming commonplace, supported by the emergence of QSCI packages such as \texttt{qiskit-addon-sqd} and \texttt{qiskit-addon-sqd-hpc}~\cite{qiskit-addon-sqd,qiskit-addon-sqd-hpc}, which IBM and its collaborators are actively developing.

In practice, however, QSCI is still restricted to an active space, i.e., a small subset of chemically relevant molecular orbitals and electrons. Consequently, electron correlation associated with orbitals outside the active space is not captured by QSCI. To recover the missing correlation and enable quantitative calculations, QSCI can be combined with classical post-processing methods. Examples include auxiliary-field quantum Monte Carlo (AFQMC) \cite{zhangQuantumMonteCarlo2003}, tailored coupled cluster (TCC) \cite{kinoshitaCoupledclusterMethodTailored2005}, and quasi-degenerate perturbation theory with general multiconfiguration reference functions (GMC-QDPT) \cite{nakanoQuasidegeneratePerturbationTheory2002}, leading to QSCI-AFQMC \cite{yoshidaAuxiliaryfieldQuantumMonte2025, danilovEnhancingAccuracyEfficiency2025}, QSCI-TCC \cite{erhartCoupledClusterMethod2025}, and QSCI-PT \cite{shiraiEnhancingAccuracyQuantumSelected2025}, respectively. The success of this approach has been demonstrated for multi-bond dissociation in small molecules such as \ce{H2O} and \ce{N2}, with QSCI-AFQMC achieving agreement with high-level benchmarks within \SI{1}{kcal/mol} \cite{yoshidaAuxiliaryfieldQuantumMonte2025}. Recent developments further show that the efficiency and scalability of QSCI with classical post-processing can be improved by incorporating doubly occupied configuration interaction (DOCI), as exemplified with DOCI-QSCI-AFQMC \cite{yoshidaDoublingSizeQuantum2026}.

Nevertheless, a major limitation of conventional classical post-processing methods is the steep growth in computational cost with system size. For example, TCC is based on coupled cluster with singles and doubles (CCSD) or on coupled cluster with singles, doubles, and perturbative triples (CCSD(T)) \cite{raghavachariFifthOrderPerturbation1989}. Since these methods scale as $O(N^6)$ or $O(N^7)$ with the number of basis functions $N$, respectively, direct canonical calculations are restricted to small system sizes even with steady increases in computational power. Consequently, applying high-level post-processing (i.e., a high-level wave-function method) directly to large systems with an explicit surrounding environment is often computationally prohibitive. Such calculations are therefore commonly limited to gas-phase/implicit-solvent models or truncated fragments containing the chemically critical region, even though environmental effects can be essential in catalysis, materials science, and drug discovery.  Reduced-scaling local-correlation variants such as DLPNO-CCSD(T) extend the reach of coupled-cluster theory \cite{riplingerEfficientApproximateCoupledcluster2013,riplingerNaturalTripleExcitations2013}.
Nonetheless, to our knowledge, no efficient open-source implementation is currently available for open-shell local-correlation coupled-cluster methods, and such methods still involve substantially higher computational cost than Kohn-Sham density functional theory (DFT). 

When the chemically demanding region is spatially localized, embedding provides a natural route to combine the accuracy of wave-function methods with the scalability of DFT. In such approaches, the region of interest is treated at the wave-function level, while the surrounding environment is described with a lower-cost method. Related embedding frameworks include density matrix embedding theory (DMET) \cite{kniziaDensityMatrixEmbedding2012,kniziaDensityMatrixEmbedding2013} and subsystem or frozen-density DFT approaches \cite{jacobSubsystemDensityFunctional2014,wesolowskiFrozenDensityEmbedding2015}. 
In this work, we use projection-based wave-function-in-DFT (WF-in-DFT) embedding \cite{manbySimpleExactDensityFunctionalTheory2012, leeProjectionBasedWavefunctioninDFTEmbedding2019}. A key advantage of this formulation is in its simplicity and that, in the DFT-in-DFT limit, it reproduces the corresponding full-system KS-DFT result when the same functional and a consistent orbital partition are used.

Embedding is therefore well suited to the present problem: it can retain an explicit DFT description of the large surrounding environment while restricting the expensive QSCI and post-processing calculations to a chemically important subsystem. Although quantum algorithms have been already combined with embedding schemes in several recent studies \cite{vorwerkQuantumEmbeddingTheories2022,shajanQuantumcentricSimulationsExtended2024,battagliaGeneralFrameworkActive2024,bickleyExtendingQuantumComputing2025,yamamotoQuantumHPCHybridComputation2026,merzCrossing12000atomBarrier2026}, most applications have remained at the level of proof-of-concept demonstrations or have not addressed the combined requirements of active-space quantum treatment, recovery of dynamical correlation, and explicit large-scale environments. It therefore remains unclear whether quantum--classical embedding strategies can deliver quantitatively reliable results for chemically realistic systems across diverse chemical environments.

In this work, we address this gap by establishing a framework that integrates QSCI, classical post-processing, and projection-based WF-in-DFT embedding. QSCI targets the chemically relevant active space, classical post-processing recovers electron correlation beyond that active space, and WF-in-DFT embedding accounts for the surrounding environment at the DFT level. We apply this framework to organic, metal-organic, and metallic systems, demonstrating its accuracy, scalability, and robustness across chemically diverse environments.

\section{Methods}

\subsection{DFT-embedded QSCI with classical post-processing}

\begin{figure*}
    \includegraphics{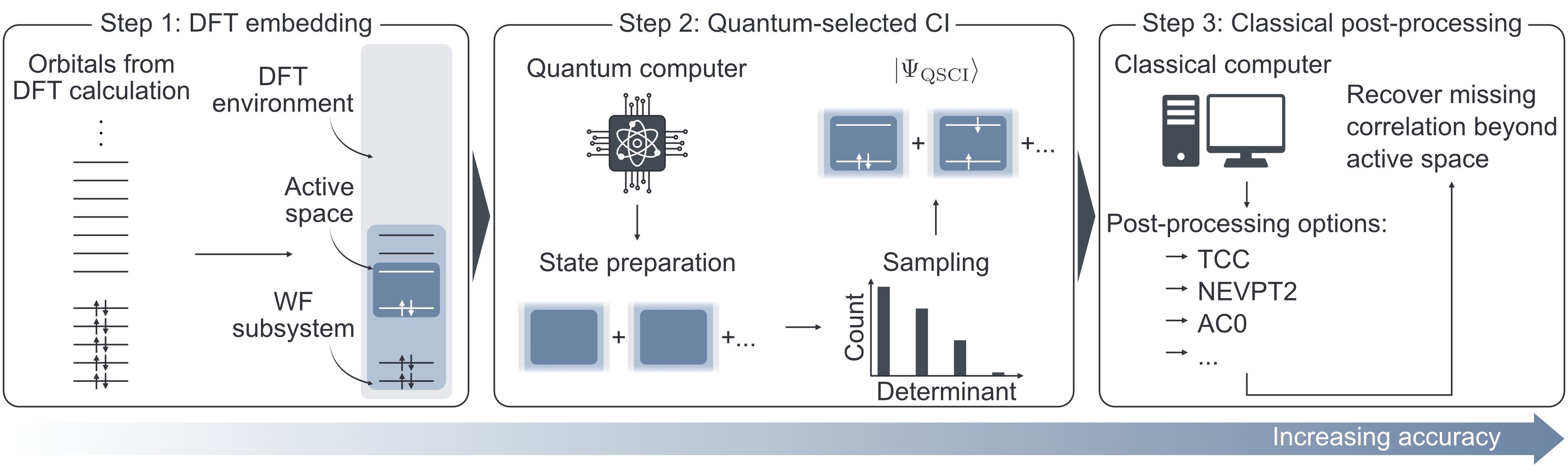}
    \centering
    \caption{\textbf{Overview of the multilevel framework combining DFT embedding, quantum computing, and classical post-processing.} 
    Step~1: Projection-based WF-in-DFT embedding partitions the molecular orbitals of the full system into a WF subsystem (light blue) and a DFT environment (gray). Step~2: An active space is selected from the WF subsystem (dark blue), and a targeted active-space quantum state is prepared on a quantum computer and sampled in the computational basis. The QSCI method is then used to approximately reconstruct the quantum state on a classical computer by diagonalizing the Hamiltonian within the determinant subspace defined using the sampled bitstrings. Step~3: Missing electron correlation is approximately recovered via classical post-processing such as TCC, NEVPT2, or AC0, enabling quantitative electronic-structure calculations beyond the active space approximation.}
    \label{fig:Method_Overview}
\end{figure*}

In this section, we describe our DFT-embedded, post-processed QSCI framework and its individual components. Fig.~\ref{fig:Method_Overview} provides an overview of the workflow, which comprises three main steps: projection-based WF-in-DFT embedding, QSCI, and classical post-processing.
We first provide a brief outline of each step and how they connect before describing each component in more detail in the subsequent sections.

In the DFT embedding step, a KS-DFT calculation on the full system is first performed to obtain the molecular orbitals. The resulting orbitals are then localized and partitioned into an active WF subsystem (light blue) and a DFT environment (gray) using projection-based WF-in-DFT embedding. This step defines an embedded Hamiltonian for the WF subsystem that incorporates the influence of the surrounding environment at the DFT level. The embedded Hamiltonian provides the starting point for accurate quantum--classical hybrid calculations of large-scale systems on quantum hardware.

In the subsequent step, a small subset of orbitals and electrons is selected from the embedded WF subsystem to define an active space (dark blue), and the resulting active-space Hamiltonian is solved on a quantum computer. The quantum device is used to prepare a state that approximates a targeted quantum state of this active space Hamiltonian. This preparation is not tied to any single algorithm. It may rely on quantum phase estimation (QPE) \cite{lloydUniversalQuantumSimulators1996,aspuru-guzikSimulatedQuantumComputation2005}, a quantum power method \cite{khinevichQuantumPowerIteration2025}, or on a variational scheme such as the VQE. From the prepared state, QSCI reconstructs the wave function classically by sampling in the computational basis to collect the most important electronic configurations and then diagonalizing the Hamiltonian within this selected subspace. The resulting compact wave function, together with the reduced density matrices derived from it, provides the reference that links the quantum solution to  a quantitative wave-function method in the post-processing step on a classical computer.

While a quantum algorithm may capture the main contributions to the static correlation, it misses electron correlation outside the active space. To approximately recover the missing electron correlation, classical post-processing is performed on the QSCI reference in the final step. In this work, we use three post-processing methods: TCC, second-order $N$-electron valence state perturbation theory (NEVPT2) \cite{angeliIntroductionNelectronValence2001,angeliNelectronValenceState2001,angeliNelectronValenceState2002}, and the AC0 correction, derived within the adiabatic connection (AC) formalism from the linearized AC-integrand approximation \cite{pastorczakCorrelationEnergyAdiabatic2018}. These methods are conceptually distinct and offer different trade-offs between accuracy and computational cost. Other post-processing methods, such as AFQMC, are equally applicable but are not explored in this work.

After completing these three steps, the final energy is obtained as the sum of the embedded QSCI energy and a post-processing correction. Following this overview, we now describe the individual components in more detail. Note that the active-space contribution to the final energy need not be the QSCI energy. As in our earlier works ~\cite{erhartCoupledClusterMethod2024,erhartCoupledClusterMethod2025,khinevichEnhancingQuantumComputations2025}, the energy from the state-preparation step, such as the QPE energy, can instead be combined with the post-processing correction in an ONIOM-like manner \cite{svenssonONIOMMultilayeredIntegrated1996}.

\subsubsection{Projection-based WF-in-DFT embedding}
In DFT, the energy of a system is expressed as a functional of its total electron density $\rho^\mathrm{total}$.
When the electron density of the total system is divided into the density $\rho^\mathrm{A}$ of an active subsystem A and the density $\rho^\mathrm{B}$ of an environment B,
\begin{equation}
    \rho^\mathrm{total} = \rho^\mathrm{A} + \rho^\mathrm{B} \,,
\end{equation}
the total energy $E[\rho^{\mathrm{total}}]$ can be correspondingly partitioned as
\begin{equation}
    E[\rho^{\mathrm{total}}] = E[\rho^\mathrm{A}] + E[\rho^\mathrm{B}] + \delta E[\rho^\mathrm{A}, \rho^\mathrm{B}] \,.
\end{equation}
$E[\rho^\mathrm{A}]$ and $E[\rho^\mathrm{B}]$ denote the energies of the isolated subsystems A and B, respectively, and $\delta E[\rho^\mathrm{A}, \rho^\mathrm{B}]$ accounts for the nonadditivity arising from the kinetic, Coulomb, and exchange-correlation terms. Among these, the contribution from the kinetic energy is the most challenging to compute. While various approximate methods have been developed, its explicit evaluation can be avoided using projection-based embedding, as outlined below \cite{manbySimpleExactDensityFunctionalTheory2012, leeProjectionBasedWavefunctioninDFTEmbedding2019}.

When a KS-DFT calculation is performed on the total system, it yields a set of orbitals and occupation numbers from which the one-particle density matrix $\boldsymbol{\gamma}^\mathrm{total}$ is constructed. By assigning subsets of these orbitals to the active subsystem A and the environment B, the total one-particle density matrix is partitioned into the density matrices $\boldsymbol{\gamma}^\mathrm{A}$ and $\boldsymbol{\gamma}^\mathrm{B}$ of the two subsystems,
\begin{equation}
    \boldsymbol{\gamma}^\mathrm{total} = \boldsymbol{\gamma}^\mathrm{A} + \boldsymbol{\gamma}^\mathrm{B},
\end{equation}
which likewise implies a decomposition of the electron density. The energy of A embedded in B is then given by
\begin{equation}
E[\boldsymbol{\gamma}^\mathrm{A}, \boldsymbol{\gamma}^\mathrm{B}] = \mathrm{tr}(\boldsymbol{\gamma}^\mathrm{A} + \boldsymbol{\gamma}^\mathrm{B})\mathbf{h} + J[\boldsymbol{\gamma}^\mathrm{A} + \boldsymbol{\gamma}^\mathrm{B}] + E_{\mathrm{xc}}[\boldsymbol{\gamma}^\mathrm{A} + \boldsymbol{\gamma}^\mathrm{B}] \,,
\end{equation}
where $\mathbf{h}$ is the core Hamiltonian in the atomic orbital (AO) basis, accounting for the kinetic energy and external potentials, and the last two terms represent the Coulomb and exchange-correlation contributions, respectively.

The contribution from the kinetic energy is eliminated in the projection-based embedding scheme by orthogonalizing the molecular orbitals of the subsystems. This is accomplished by introducing a level-shift operator
\begin{equation}
    \mu \mathbf{P}^\mathrm{B} = \mu \mathbf{S} \boldsymbol{\gamma}^\mathrm{B} \mathbf{S}
\end{equation}
with a positive real number $\mu$ and the overlap matrix $\mathbf{S}$ in the AO basis. If $\mu$ is chosen sufficiently large, the operator $\mu \mathbf{P}^\mathrm{B}$ shifts the energies of the orbitals in the environment B high enough that they no longer hybridize with those in the active subsystem A, and the two sets of orbitals become mutually orthogonal.

By incorporating the level-shift operator into the core Hamiltonian of the embedded subsystem A, one obtains
\begin{equation}
    \begin{aligned}
        \mathbf{h}^{\text{A in B}} &= \mathbf{h} + \mathbf{J}[\boldsymbol{\gamma}^\mathrm{A} + \boldsymbol{\gamma}^\mathrm{B}] - \mathbf{J}[\boldsymbol{\gamma}^\mathrm{A}] + \mathbf{v}_{\mathrm{xc}}[\boldsymbol{\gamma}^\mathrm{A} + \boldsymbol{\gamma}^\mathrm{B}]\\
    & \quad- \mathbf{v}_{\mathrm{xc}}[\boldsymbol{\gamma}^\mathrm{A}] + \mu \mathbf{P}^\mathrm{B} \,.
    \end{aligned}
\end{equation}
Here, $\mathbf{J}$ and $\mathbf{v}_\mathrm{xc}$ denote the Coulomb and exchange-correlation potentials in the AO basis, respectively. Using $\mathbf{h}^{\text{A in B}}$ as the effective one-electron Hamiltonian of the active subsystem A in the presence of the environment B, a self-consistent field (SCF) calculation is performed to obtain the reference orbitals for the subsequent QSCI and classical post-processing steps. The corresponding embedded Fock matrix is defined as
\begin{equation}
\mathbf{F}^{\mathrm{A}}
=
\mathbf{h}^{\mathrm{A\ in\ B}}
+
\mathbf{g}[\boldsymbol{\gamma}^{\mathrm{A}}] \,,
\end{equation}
where $\mathbf{g}[\boldsymbol{\gamma}^{\mathrm{A}}]$ contains the usual two-electron terms for the SCF calculation.

Once the wave function $\Psi^\text{A}$ for the embedded subsystem A has been obtained, the total energy of the resulting WF-in-DFT system is given by
\begin{equation}
\begin{aligned}
E[\Psi^\text{A}; \boldsymbol{\gamma}^\text{B}] &= 
\langle \Psi^\text{A} | \hat{H}^\text{A in B} | \Psi^\text{A} \rangle \\
&\quad - \mathrm{tr}\, \boldsymbol{\gamma}^\text{A} \left( \mathbf{v}_{\mathrm{xc}}[\boldsymbol{\gamma}^\text{A} + \boldsymbol{\gamma}^\text{B}] - \mathbf{v}_{\mathrm{xc}}[\boldsymbol{\gamma}^\text{A}] \right) \\
&\quad + E_{\mathrm{xc}}[\boldsymbol{\gamma}^\text{A} + \boldsymbol{\gamma}^\text{B}] - E_{\mathrm{xc}}[\boldsymbol{\gamma}^\text{A}] + E[0; \boldsymbol{\gamma}^\text{B}] \,.
\end{aligned}
\end{equation}
$\hat{H}^\text{A in B}$ denotes the corresponding embedded Hamiltonian, whose one-electron part is defined by $\mathbf{h}^{\text{A in B}}$.

\subsubsection{Quantum-selected configuration interaction (QSCI)}
In the full configuration interaction (FCI) method, the wave function is written as a linear combination of all Slater determinants that can be formed from the chosen orbital space,
\begin{equation}
    \left| \Psi_{\mathrm{FCI}} \right\rangle = \sum_{i=1}^{N} c_i \left| \Phi_i \right\rangle \,,
\end{equation}
where $\left| \Phi_i \right\rangle$ is the $i$th Slater determinant, $c_i$ the corresponding CI coefficient, and $N$ the total number of determinants. FCI yields the exact ground state within the chosen orbital basis but is intractable for large systems due to its combinatorially increasing computational cost. For practical applications, it is therefore necessary to truncate the Fock space. The QSCI method addresses this by identifying the most important states and using them to construct the wave function. 
For this, a trial wave function that approximates the state of interest is prepared on a quantum computer. The state is sampled by performing $N_\mathrm{shot}$ projective measurements in the computational basis, which yields a distribution of bitstrings that reflects their probabilities in the wave function. The $R$ most frequently observed bitstrings define the set of basis states $\left\{ \left| \Phi_i \right\rangle \right\}_{i=1}^{R}$, which is then used to construct the QSCI wave function
\begin{equation}
    \left| \Psi_{\mathrm{QSCI}} \right\rangle = \sum_{i=1}^{R} c_i \left| \Phi_i \right\rangle \,.
    \label{eq:Psi_QSCI}
\end{equation}
To obtain the corresponding set of coefficients $\left\{ c_i \right\}$, the $R \times R$ effective Hamiltonian $\mathbf{H}^{\mathrm{eff}}$ is defined in the subspace spanned by the $R$ selected configurations, with matrix elements
\begin{equation}
    H^{\mathrm{eff}}_{ij} = \langle \Phi_i | \hat{H} | \Phi_j \rangle \,,
\end{equation}
where $\hat H$ is the electronic Hamiltonian. Solving the subspace eigenvalue problem
\begin{equation}
    \mathbf{H}^{\mathrm{eff}} \mathbf{c} = E \mathbf{c} \,
\end{equation}
yields the energy eigenvalue $E$ and the coefficient vector $\mathbf{c}$ of the ground state. In our workflow, $\hat H$ is taken as the embedded Hamiltonian $\hat{H}^\text{A in B}$ obtained from the projection-based WF-in-DFT embedding.

Finite-shot sampling can introduce an imbalance between $\alpha$ and $\beta$ spins in the sampled wave function. To preserve the target spin-projection sector specified by $(N_\alpha,N_\beta)$, each sampled state is first written as
\begin{equation}
    |\Phi_i\rangle = |\Phi_i^{(\alpha)}\rangle |\Phi_i^{(\beta)}\rangle \,.
    \label{eq:spin}
\end{equation}
The determinant subspace used in QSCI is then formed by taking the Cartesian product of bitstrings for each spin, i.e.,
\begin{equation}
    \big\{\,|\tilde{\Phi}_i\rangle\,\big\}
= \big\{\, |\Phi_m^{(\alpha)}\rangle |\Phi_n^{(\beta)}\rangle
\;\big|\; 1\le m\le R,\; 1\le n\le R \,\big\}.
\end{equation}
While the enlarged subspace spanned by $\big\{\,|\tilde{\Phi}_i\rangle\,\big\}$ increases the dimension of $\mathbf{H}^{\mathrm{eff}}$, it also raises the likelihood of capturing relevant determinants, which has been shown to improve accuracy of QSCI~\cite{yoshidaAuxiliaryfieldQuantumMonte2025,robledo-morenoChemistryScaleExact2025}.

\subsubsection{Post-processing using TCC}

In the conventional coupled-cluster (CC) method, the wave function is written as 
\begin{equation}
    \lvert \Psi_{\mathrm{CC}} \rangle
= e^{\hat{T}} \, \lvert \Psi_0 \rangle \,,
\end{equation}
where $\lvert \Psi_0 \rangle$ is the reference Slater determinant and $\hat{T}$ the cluster operator.
For single-reference systems near equilibrium, CCSD(T) is widely regarded as the gold standard because it provides a highly accurate treatment of dynamical electron correlation. However, CCSD(T) becomes unreliable when electron correlation is strong. This limitation can be mitigated by incorporating an active-space CI calculation through the TCC method \cite{kinoshitaCoupledclusterMethodTailored2005}. 
In the TCC framework, the cluster operator $\hat{T}$ is partitioned into two parts to approximately separate dynamical and static electron correlation:
\begin{equation}
    \lvert \Psi_{\mathrm{TCC}} \rangle
= e^{\hat{T}^{\mathrm{external}}}  e^{\hat{T}^{\mathrm{active}}} \lvert \Psi_0 \rangle \,.
\end{equation}
$\hat{T}^{\mathrm{active}}$ acts solely within the active space and serves to capture most of the static correlation. $\hat{T}^{\mathrm{external}}$ contains all other excitations and accounts for the remaining dynamical correlation in the full system. 
In the TCCSD ansatz, $\hat{T}^{\mathrm{active}}$ and $\hat{T}^{\mathrm{external}}$ are given by
\begin{equation}
    \begin{aligned}
        \hat{T}^{\mathrm{active}}
&= \hat{T}^{\mathrm{active}}_{1} + \hat{T}^{\mathrm{active}}_{2} \\
&= \sum_{i',a'} t_{i'}^{a'}\,\hat{a}^{\dagger}_{a'}\hat{a}_{i'}
 \;+\frac{1}{4}\; \sum_{i',j',a',b'} t_{i'j'}^{a'b'}\,\hat{a}^{\dagger}_{a'}\hat{a}^{\dagger}_{b'}\hat{a}_{j'}\hat{a}_{i'}, \\
&\qquad i',j',a',b' \in \text{active space}
    \end{aligned}
\end{equation}
and
\begin{equation}
    \begin{aligned}
        \hat{T}^{\mathrm{external}}
&= \hat{T}^{\mathrm{external}}_{1} + \hat{T}^{\mathrm{external}}_{2} \\
&= \sum_{i,a} t_i^{a}\,\hat{a}^{\dagger}_{a}\hat{a}_{i}
 \;+\frac{1}{4}\; \sum_{i,j,a,b} t_{ij}^{ab}\,\hat{a}^{\dagger}_{a}\hat{a}^{\dagger}_{b}\hat{a}_{j}\hat{a}_{i}, \\
&\qquad \{i,j,a,b\} \not\subset \text{active space}.
    \end{aligned}
\end{equation}
Here, $t_i^{a}$, $t_{i'}^{a'}$, $t_{ij}^{ab}$, and $t_{i'j'}^{a'b'}$ are the coupled-cluster single- and double-excitation amplitudes. $\hat{a}^{\dagger}_p$ and $\hat{a}_p$ denote the creation and annihilation operators acting on spin-orbital $p$, respectively. Indices $i$, $i'$, $j$, and $j'$ run over occupied orbitals of the reference wave function, whereas $a$, $a'$, $b$, and $b'$ run over unoccupied orbitals. Note that in $\hat{T}^{\mathrm{active}}$ all indices are restricted to the active space, whereas in $\hat{T}^{\mathrm{external}}$ at least one index lies outside of it.

To embed the static correlation recovered by the active-space CI calculation into the CC wave function, the cluster operator is \enquote{tailored} by constraining the amplitudes in $\hat{T}^{\mathrm{active}}$ to reproduce the active-space CI expansion. For the TCCSD ansatz, this yields
\begin{align}
    \hat{T}^{\mathrm{active}}_{1} &= \hat{C}_{1}, \\
\hat{T}^{\mathrm{active}}_{2} &= \hat{C}_{2} - \tfrac{1}{2}\,\hat{C}_{1}^{2}\,,
\end{align}
where $\hat{C}_{1}$ and $\hat{C}_{2}$ denote the CI single- and double-excitation operators, respectively. In this work, $\hat{C}_{1}$ and $\hat{C}_{2}$ are reconstructed from the QSCI wave function eqn.~(\ref{eq:Psi_QSCI}), retaining only the most important excitations \cite{erhartCoupledClusterMethod2025}. Dynamical correlation is then recovered by solving the standard coupled-cluster equations for $\hat{T}^{\mathrm{external}}$ while keeping $\hat{T}^{\mathrm{active}}$ fixed.

The accuracy can be further improved by incorporating the perturbative triples correction, resulting in TCCSD(T). To prevent double-counting, all active single and double amplitudes are set to zero during the evaluation of the perturbative triples correction~\cite{lyakhTailoredCCSDTDescription2011}.

\subsubsection{Post-processing using NEVPT2}
NEVPT2 is a multireference perturbation theory for recovering dynamical correlation from an active-space reference \cite{angeliIntroductionNelectronValence2001,angeliNelectronValenceState2001,angeliNelectronValenceState2002}.
Here, it is applied as a post-processing step on top of the QSCI wave function, using the strongly contracted variant. 

As a perturbation theory, NEVPT2 is formulated by partitioning the Hamiltonian as
\begin{equation}
    \hat{H}=\hat{H}^{(0)}+\hat{V},
\end{equation}
where $\hat{H}^{(0)}$ is the zeroth-order Hamiltonian and $\hat{V}=\hat{H}-\hat{H}^{(0)}$ is the perturbation. To ensure rapid convergence, $\hat{H}^{(0)}$ is chosen to approximate $\hat{H}$ closely, making $\hat{V}$ small.

A central aspect of NEVPT2 is choosing $\hat{H}^{(0)}$ as the Dyall Hamiltonian \cite{dyallChoiceZerothorderHamiltonian1995},
\begin{equation}
\hat{H}^{(0)} = \hat{H}_\text{active} + \sum_i\epsilon_i a_i^\dagger a_i + \sum_a\epsilon_a a_a^\dagger a_a + C \,,
\end{equation}
where $\hat{H}_\text{active}$ contains all terms that involve the active space. The summations over $i$ and $a$ run over doubly occupied (core) and empty (virtual) orbitals, respectively. $\epsilon_i$ and $\epsilon_a$ are semicanonical orbital energies for the core and virtual orbitals, respectively. $C$ is a constant obtained from contributions of the core orbitals.

With this choice of $\hat{H}^{(0)}$ and the QSCI wave function (eqn.~(\ref{eq:Psi_QSCI})) as the reference state, the NEVPT2 energy correction is obtained from a second-order expansion in $\hat{V}$:
\begin{equation}
    E^{(2)}=\langle \Psi_{\mathrm{QSCI}}|\hat V\,\hat Q\,(E_0^{(0)}-\hat H^{(0)})^{-1}\hat Q\,\hat V|\Psi_{\mathrm{QSCI}}\rangle .
\end{equation}
Here, $\hat Q$ projects onto the space outside the QSCI active-space reference and $E_0^{(0)}$ is the zeroth-order energy obtained from $\hat H^{(0)}$.
For the QSCI reference, the second-order correction involves intermediate configurations generated by one- and two-electron transfers between the active space and the core and virtual orbital spaces. 

\subsubsection{Post-processing using AC0}

AC0 is based on an adiabatic connection (AC) that interpolates between a chosen reference Hamiltonian and the fully interacting electronic Hamiltonian \cite{pernalElectronCorrelationAdiabatic2018}. Compared to representative multireference perturbation methods such as CASPT2 and NEVPT2, AC0 does not require reduced density matrices higher than second order, making it computationally less demanding for recovering dynamical correlation outside the active space \cite{pastorczakCorrelationEnergyAdiabatic2018}.

The AC Hamiltonian is defined as
\begin{equation}
    \hat{H}^\alpha = \hat{H}^{(0)} + \alpha\hat{H}'
\end{equation}
with
\begin{equation}
    	\hat{H}' = \hat{H} - \hat{H}^{(0)} \,,
\end{equation}
where $0 \leq \alpha \leq 1$ is a coupling constant and $\hat{H}^{(0)}$ is chosen such that the active-space problem is described at $\alpha=0$. In our workflow, the QSCI wave function eqn.~(\ref{eq:Psi_QSCI}) is used as the active-space reference.

When the correlation energy is defined as
\begin{equation}
    E_\mathrm{corr} = E_\mathrm{exact} - E_\mathrm{reference} \,,
\end{equation}
where $E_\mathrm{exact}$ is the exact energy and $E_\mathrm{reference}$ is the energy of the active-space reference, it can be shown that the exact AC expression for the correlation energy $E^\mathrm{AC}_\mathrm{corr}$ is given by
\begin{align}
    E^\mathrm{AC}_\mathrm{corr} = \int_0^1 \left(W^\alpha + \Delta^\alpha\right)\mathrm{d}\alpha \,.
    \label{eq:E_AC}
\end{align}
Here, $W^\alpha$ comprises contributions from two-electron interactions evaluated with the $\alpha$-dependent wave function and $\Delta^\alpha$ accounts for one-electron density changes along the path.

For practical applications, eqn.~(\ref{eq:E_AC}) can be simplified by assuming that the one-electron density matrix remains constant along the AC path, which causes $\Delta^\alpha$ to vanish, and by evaluating $W^\alpha$ within the extended random phase approximation (ERPA). In AC0, the integrand is further approximated by a first-order expansion around $\alpha=0$:
\begin{equation}
    W^\alpha \approx W^{\alpha=0} + \left.\frac{\mathrm{d}W}{\mathrm{d}\alpha}\right|_{\alpha=0}\alpha \,.
\end{equation}
It can be shown that $W^{\alpha=0}=0$ for an active-space reference wave function ~\cite{pastorczakCorrelationEnergyAdiabatic2018}. In this study, we use the QSCI wave function as an approximate active-space reference. With $W^{(1)} = \left.\frac{\mathrm{d}W}{\mathrm{d}\alpha}\right|_{\alpha=0}$, this yields
\begin{equation}
    E^\mathrm{AC}_\mathrm{corr} \approx \int_0^1 \left(W^{\alpha=0} + W^{(1)}\alpha \right)\mathrm{d}\alpha = \frac{W^{(1)}}{2} \equiv E^\mathrm{AC0}_\mathrm{corr}
\end{equation}
for the AC0 expression for the correlation energy.

\subsection{Computational details}

Projection-based WF-in-DFT embedding was performed using the commercial QSimulate-QM electronic-structure package licensed from QSimulate, which is built on the BAGEL program package \cite{shiozakiBAGELBrilliantlyAdvanced2018}. Orbitals for the active WF subsystem were selected using Pipek–Mezey localization \cite{pipekFastIntrinsicLocalization1989} followed by the atomic valence active space (AVAS) construction \cite{sayfutyarovaAutomatedConstructionMolecular2017}. The DFT functionals, basis sets, and sizes of the active WF subsystems used for each system are reported in the corresponding Results sections.

Wave functions were prepared using quantum circuits via VQE simulations carried out classically in chemqulacs \cite{Chemqulacs2023}. The hardware-efficient ansatz was used \cite{kandalaHardwareefficientVariationalQuantum2017}. For the fermion-to-qubit mapping, the symmetry-conserving Bravyi–Kitaev (SCBK) mapping \cite{bravyiTaperingQubitsSimulate2017} was employed. The circuit parameters were optimized in the classical simulations of the VQE. The optimized active-space circuits were subsequently executed and measured using a shot budget of \num{10000} on the quantum computer at the University of Osaka. The device features a 144-physical-qubit superconducting chip, but each calculation used only the subset of qubits required by the corresponding active-space mapping. This chip is similar to that in Ref.~\cite{springFastMultiplexedSuperconductingQubit2025}, but operates at qubit frequencies in the \SI{5}{GHz} band rather than the \SI{8}{GHz} band. QURI Parts version 0.20 \cite{QURIParts2024} and QURI Parts OQTOPUS \cite{OQTOPUS2025, kakukoPracticalOpenSourceSoftware2025} were employed to interface the quantum and classical computations and access the quantum device via cloud-based services. 
To account for spin symmetry when constructing the effective Hamiltonian from the sampled determinants (eqn.~(\ref{eq:spin})), the union of the $\alpha$- and $\beta$-spin configuration sets was formed by adding configurations that are present in one spin sector but absent from the other.
Active space sizes for the QSCI simulations are given in the corresponding Results sections.

SCF calculations as well as post-processing with strongly contracted NEVPT2 were conducted using PySCF version 2.2.1 \cite{sunPySCFPythonbasedSimulations2018,sunRecentDevelopmentsPySCF2020}. 
For AC0 calculations, we employed the CAS-AC0 module for PySCF developed by Quantinuum ~\cite{PySCF-AC02024}.
TCCSD(T) post-processing was performed using an in-house implementation built on PySCF \cite{erhartCoupledClusterMethod2025}.

The initial structure of ethanol was obtained via geometry optimization using ORCA version 6.1 \cite{neeseORCAProgramSystem2012, neeseORCAQuantumChemistry2020, neeseSHARKIntegralGeneration2023, neeseSoftwareUpdateORCA2025} at the PBE/cc-pVTZ level. For propionitrile, the geometry provided in Ref.~\cite{beranProjectionBasedDensityMatrix2023} was used. The geometries of the reactant, transition, and product states for the Menshutkin $\mathrm{S_N2}$ reaction in gas phase and inside a carbon nanotube were taken from Ref.~\cite{claudinoAutomaticPartitionOrbital2019}. The geometry of the HKUST-1 cluster model, including the adsorbed water molecule, was generated from the periodic unit cell provided in Ref.~\cite{xueTheoreticalInsightsInitial2020}. A closed cage was cleaved from the unit cell, and terminal carbon atoms were capped with hydrogen using a C--H bond length of \SI{1.07}{\angstrom} to minimize boundary effects. 
Geometry optimization on the water molecule was performed using the large model of the in-house trained machine learning interatomic potential MACE-Osaka24 \cite{shiotaTamingMultiDomainFidelity2024}.

The reaction path for CO hopping on the HEA nanoparticle was obtained using NEB \cite{jonssonNudgedElasticBand1998, henkelmanImprovedTangentEstimate2000} calculations performed with CP2K \cite{kuhneCP2KElectronicStructure2020}. 
The nanoparticle was constructed by randomly assigning Ir, Pd, Pt, Ru, and Rh atoms at equal ratios to a truncated octahedron structure, followed by geometry optimization. To define the initial and final states for the NEB calculation, a CO molecule was placed at the respective adsorption sites, and the resulting structures were optimized while keeping the nanoparticle geometry fixed. All calculations employed the PBE exchange-correlation functional \cite{perdewGeneralizedGradientApproximation1996}, the GTH-PBE pseudopotentials \cite{goedeckerSeparableDualspaceGaussian1996}, and the DZVP-MOLOPT-PBE-GTH basis set \cite{vandevondeleGaussianBasisSets2007}.

\section{Results and discussion}

We first validate the accuracy of our DFT-embedded, post-processed QSCI framework on bond stretching for two small molecules of increasing difficulty. 
Such benchmarks are widely used because the electronic structure develops increasing multireference character upon stretching, making single-reference methods unreliable in the dissociation region.
They are therefore well-suited for highlighting how a compact active space in QSCI captures the dominant contributions to static correlation and how DFT embedding and post-processing recover dynamical correlation.
Following these benchmarks, we demonstrate the scalability and versatility of our quantum--classical approach on three chemically distinct large-scale systems of practical relevance.

\subsection{O--H bond dissociation in ethanol}

\begin{figure}
    \includegraphics{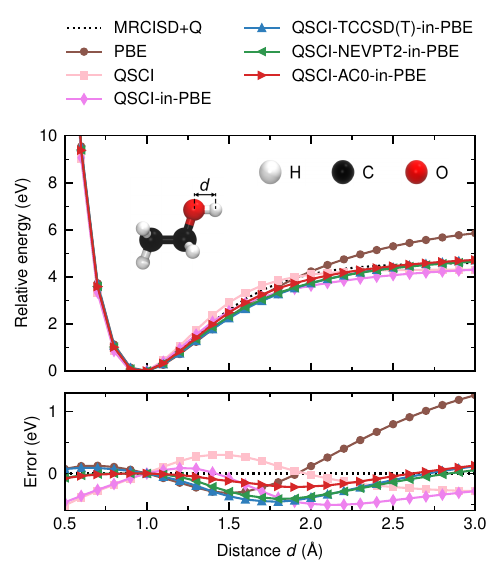}
    \centering
    \caption{\textbf{O--H bond dissociation in ethanol.} 
    Relative potential energy curves (top) for O--H bond dissociation in ethanol using the cc-pVDZ basis set.
    Corresponding errors (bottom) are defined as deviations from the MRCISD+Q reference. QSCI calculations used an active space of (2e,2o).}
    \label{fig:Ethanol_dissociation}
\end{figure}

For the first benchmark, we consider the dissociation of the O--H bond in ethanol. As a reference, multireference configuration interaction with single and double excitations including the Davidson (+Q) correction (MRCISD+Q) calculations were carried out using ORCA version 6.1 \cite{neeseORCAProgramSystem2012, neeseORCAQuantumChemistry2020, neeseSHARKIntegralGeneration2023, neeseSoftwareUpdateORCA2025}.
Fig.~\ref{fig:Ethanol_dissociation} shows the relative potential energy curves (top panel) and the corresponding errors (bottom panel), defined as deviations from the MRCISD+Q reference calculated with an active space of (8e,6o). 
The active WF subsystem comprises the electrons and orbitals on the O and H atoms involved in the O--H bond, while the remainder of the molecule constitutes the DFT environment described at the PBE level. 

All methods reproduce the position of the minimum at \SI{1.0}{\angstrom} and follow the MRCISD+Q reference curve within \SI{0.5}{eV} for distances up to $\sim$\SI{2.1}{\angstrom}. For larger distances, PBE diverges from the reference and increasingly overestimates the relative energy. This reflects that PBE treats dynamical correlation near the equilibrium distance reasonably well, but struggles with static correlation as the multireference character of the electronic structure grows in the dissociation region.

In contrast, plain QSCI follows the MRCISD+Q reference more closely than PBE in the dissociation region, but shows larger deviations for distances below \SI{2}{\angstrom}. 
The improved agreement at larger distances can be attributed to the multideterminant ansatz of the QSCI wave function, which captures a large portion of the static correlation. At shorter distances, however, the restricted active space misses the dominant dynamical correlation, which arises largely from excitations outside the active space. 

Embedding QSCI into a PBE environment (QSCI-in-PBE) reduces the overestimation from the minimum up to $\sim$\SI{1.5}{\angstrom}, while preserving the improved description at larger distances. Post-processing QSCI-in-PBE using TCCSD(T), NEVPT2, and AC0 further improves the description. Although the relative energy between \SI{1.0}{\angstrom} and \SI{2.5}{\angstrom} is still underestimated by up to \SI{0.5}{eV}, the error decreases at larger distances and lies between \SI{0.05}{eV} and \SI{0.1}{eV} at \SI{3.0}{\angstrom}. Note that AC0 achieves the most balanced description across the entire region by reducing the maximum deviation between \SI{1.0}{\angstrom} and \SI{2.5}{\angstrom} to \SI{0.2}{eV}.

\subsection{C$\equiv$N triple-bond stretching in propionitrile}

\begin{figure}
    \includegraphics{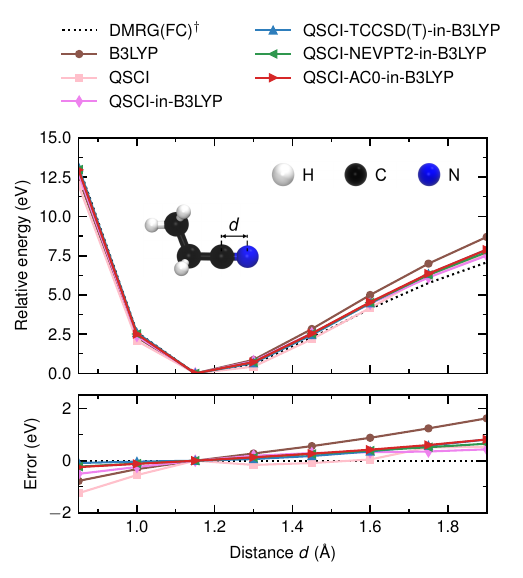}
    \centering
    \caption{\textbf{C$\equiv$N triple-bond stretching in propionitrile.} Relative potential energy curves (top) for C$\equiv$N triple-bond stretching in propionitrile using the cc-pVDZ basis set. 
    Corresponding errors (bottom) are defined as deviations from the DMRG(FC) reference. QSCI calculations used an active space of (6e,6o). $^\dagger$Values are taken from Ref. \cite{beranProjectionBasedDensityMatrix2023}.}
    \label{fig:Propionitrile_dissociation}
\end{figure}

We next examine the stretching of the C$\equiv$N triple bond in propionitrile. 
The relative potential energy curves are presented in Fig.~\ref{fig:Propionitrile_dissociation} (top panel) and the corresponding errors (bottom panel) are given as deviations from the results obtained with frozen-core (FC) density matrix renormalization group (DMRG), denoted as DMRG(FC), reported by Beran et al. \cite{beranProjectionBasedDensityMatrix2023}.
The active WF subsystem is defined by the electrons and orbitals localized on the C and N atoms of the nitrile group, while the remaining part of the molecule is assigned to the DFT environment treated at the B3LYP level. 

All methods reproduce the position of the minimum at \SI{1.15}{\angstrom} of the DMRG(FC) reference curve, but they systematically over- and underestimate the energy at distances above and below \SI{1.15}{\angstrom}, respectively.
The largest deviations are observed for B3LYP as the C$\equiv$N triple bond is stretched, consistent with the strong multireference character in the dissociation region. 

Plain QSCI shows better agreement with the DMRG(FC) reference above \SI{1.15}{\angstrom}, but has the largest errors at shorter distances. While the active space captures static correlation at larger distances, its limited size misses much of the dynamical correlation near the equilibrium geometry. 
Unlike for ethanol, embedding QSCI into a B3LYP environment (QSCI-in-B3LYP) improves the description not only in the stretched region but also at compressed bond lengths.

By post-processing QSCI-in-B3LYP with TCCSD(T), NEVPT2, and AC0, the deviation from the DMRG(FC) reference curve is further reduced below \SI{1.15}{\angstrom}. The errors are comparable to QSCI-in-B3LYP up to $\sim$\SI{1.5}{\angstrom}, but the post-processed variants show larger deviations toward dissociation. The better agreement of QSCI-in-B3LYP with the DMRG(FC) reference in this region indicates partial error cancellation between errors introduced by DFT embedding and by neglecting electron correlation outside the active space. 

The extent and benefit of error cancellation depend on the system and on the partitioning into the active WF subsystem and the DFT environment. Since the WF region can be systematically enlarged to reduce the DFT embedding error, recovering electron correlation through post-processing provides a more reliable route for obtaining accurate results. The following three large-scale applications demonstrate the scalability of DFT-embedded QSCI to realistic systems, while also showing the importance of post-processing for achieving systematically improvable accuracy and robust performance.

\subsection{Menshutkin $\mathbf{S_N2}$ reaction in a CNT}

\begin{figure}[b]
    \includegraphics[width=0.48\textwidth]{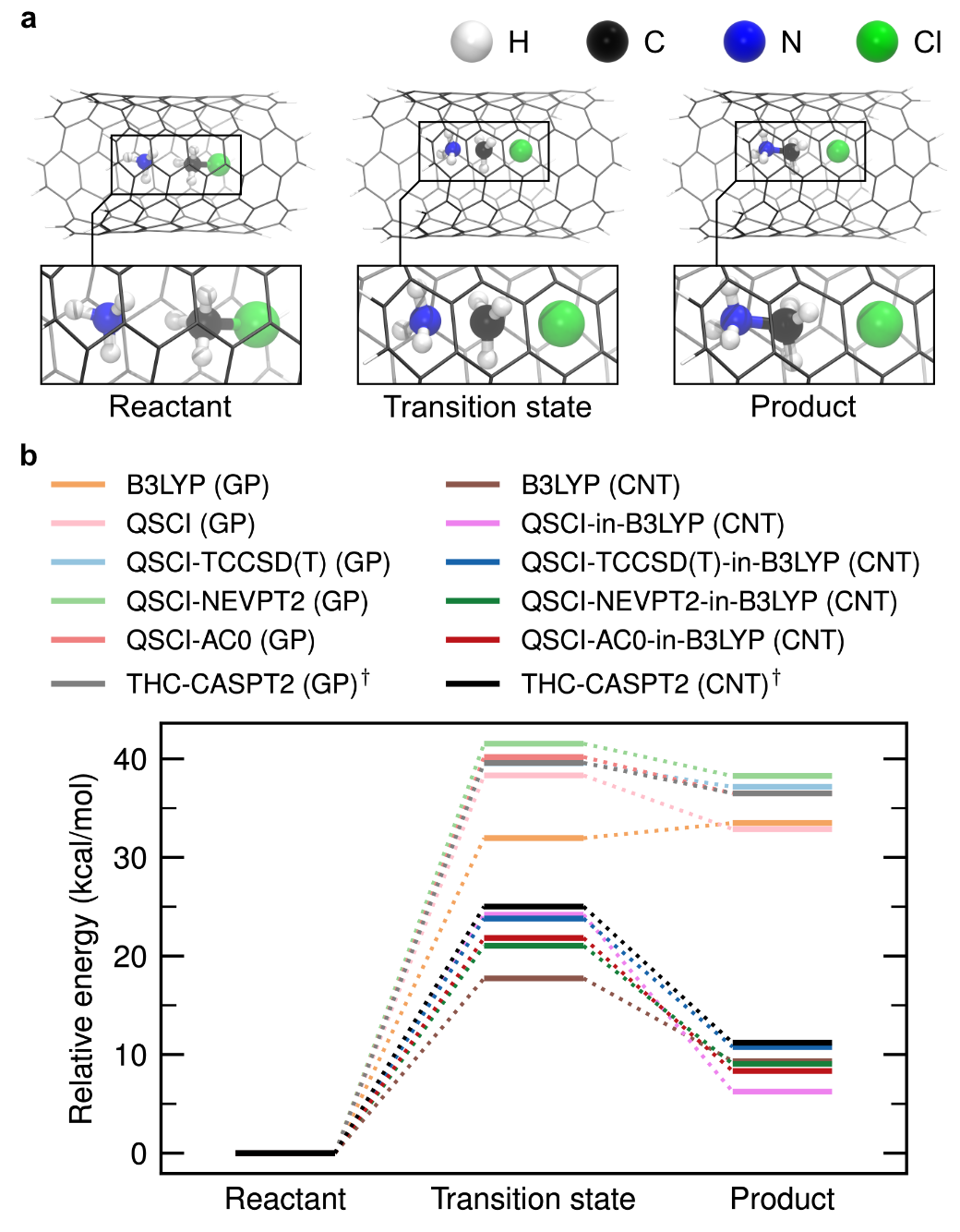}
    \centering
    \caption{\textbf{Results for the Menshutkin $\mathbf{S_N2}$ reaction in gas phase (GP) and inside a carbon nanotube (CNT).} (a) Illustration of the reactant, transition state, and product geometries of the Menshutkin $\mathrm{S_N2}$ reaction inside a CNT. (b) Energy profiles along the reaction path relative to the reactant energy computed with the 6-31G* basis set. QSCI calculations used an active space of (4e,4o). $^\dagger$Values are taken from Refs.~\cite{songReducedScalingCASPT22018, claudinoAutomaticPartitionOrbital2019}.}
    \label{fig:CNT_results}
\end{figure}

For our first demonstration on large-scale systems, we investigate the Menshutkin $\mathrm{S_N2}$ reaction inside a CNT, in which ammonia and chloromethane react to form methylammonium and chloride ions \cite{nikolaimenschutkinBeitragenZurKenntnis1890,nikolaimenschutkinUberAffinitatskoeffizientenAlkylhaloide1890}.
This reaction involves charge separation along the pathway, which makes the corresponding energy profile highly sensitive to the surrounding dielectric environment \cite{amovilliMCSCFStudySN21998}.
Computational studies suggest that confinement inside a CNT can lower the activation barrier relative to the gas phase (GP) \cite{hallsChemistryCarbonNanotubes2002, giacintoCNTConfinementEffectsMenshutkin2016, songReducedScalingCASPT22018, claudinoAutomaticPartitionOrbital2019}.
In the following, we show that our DFT-embedded, post-processed QSCI framework quantitatively captures this confinement effect.

Fig.~\ref{fig:CNT_results}a shows the reactant, transition state, and product of the Menshutkin $\mathrm{S_N2}$ reaction inside the CNT. The corresponding energies relative to the reactant are shown in Fig.~\ref{fig:CNT_results}b.
The active WF subsystem comprises all 36 electrons of the reacting molecules, while the entire CNT is treated as the DFT environment described at the B3LYP level. 
We compare our results with the CASPT2 calculations from Ref.~\cite{songReducedScalingCASPT22018}, which employ a reduced-scaling formulation based on the supporting subspace technique and tensor hyper-contraction (THC), denoted as THC-CASPT2, and use an active space of (4e,4o).

For the GP calculations, B3LYP yields an incorrect energy profile, underestimating the transition-state energy and placing it below the product, consistent with Ref.~\cite{claudinoAutomaticPartitionOrbital2019}.
Plain QSCI reproduces the energy profile qualitatively, but may underestimate the reaction energy compared to the THC-CASPT2 reference. Once QSCI is post-processed with TCC, NEVPT2, or AC0, the resulting energy profiles agree with the reference within \SI{2}{kcal/mol}.

When the reaction is confined inside the CNT, B3LYP reproduces the energy profile along the reaction coordinate qualitatively, but underestimates the barrier height by \SI{7}{kcal/mol} compared to the THC-CASPT2 reference. By contrast, QSCI-in-B3LYP matches the reference barrier height within \SI{1}{kcal/mol} but underestimates the relative energy of the product by \SI{5}{kcal/mol}. This trend mirrors the GP calculations, suggesting that the dominant correlation effects governing the barrier height can be captured with the compact active space, whereas the product requires additional correlation beyond the active space.

By restoring dynamical correlation through the TCC post-processing step, the QSCI-TCCSD(T)-in-B3LYP energy profile is in quantitative agreement with the THC-CASPT2 reference, with deviations of $\sim$\SI{1}{kcal/mol} at both the transition state and product. Post-processing with NEVPT2 and AC0, however, lowers the relative energies of the transition state and product by $\sim$\SI{4}{kcal/mol} and $\sim$\SI{3}{kcal/mol}, respectively, compared to the THC-CASPT2 reference. Still, both methods improve the barrier height relative to B3LYP and the reaction energy relative to QSCI-in-B3LYP. Note that within each post-processing method, the relative energies of the transition state and product are shifted by roughly the same amount. In contrast, the deviations in QSCI-in-B3LYP are state-dependent, suggesting that the apparent agreement of the barrier height may reflect a fortunate instance of error cancellation.

\subsection{Water adsorption on an HKUST-1 MOF cluster}

\begin{figure}
    \includegraphics[width=0.48\textwidth]{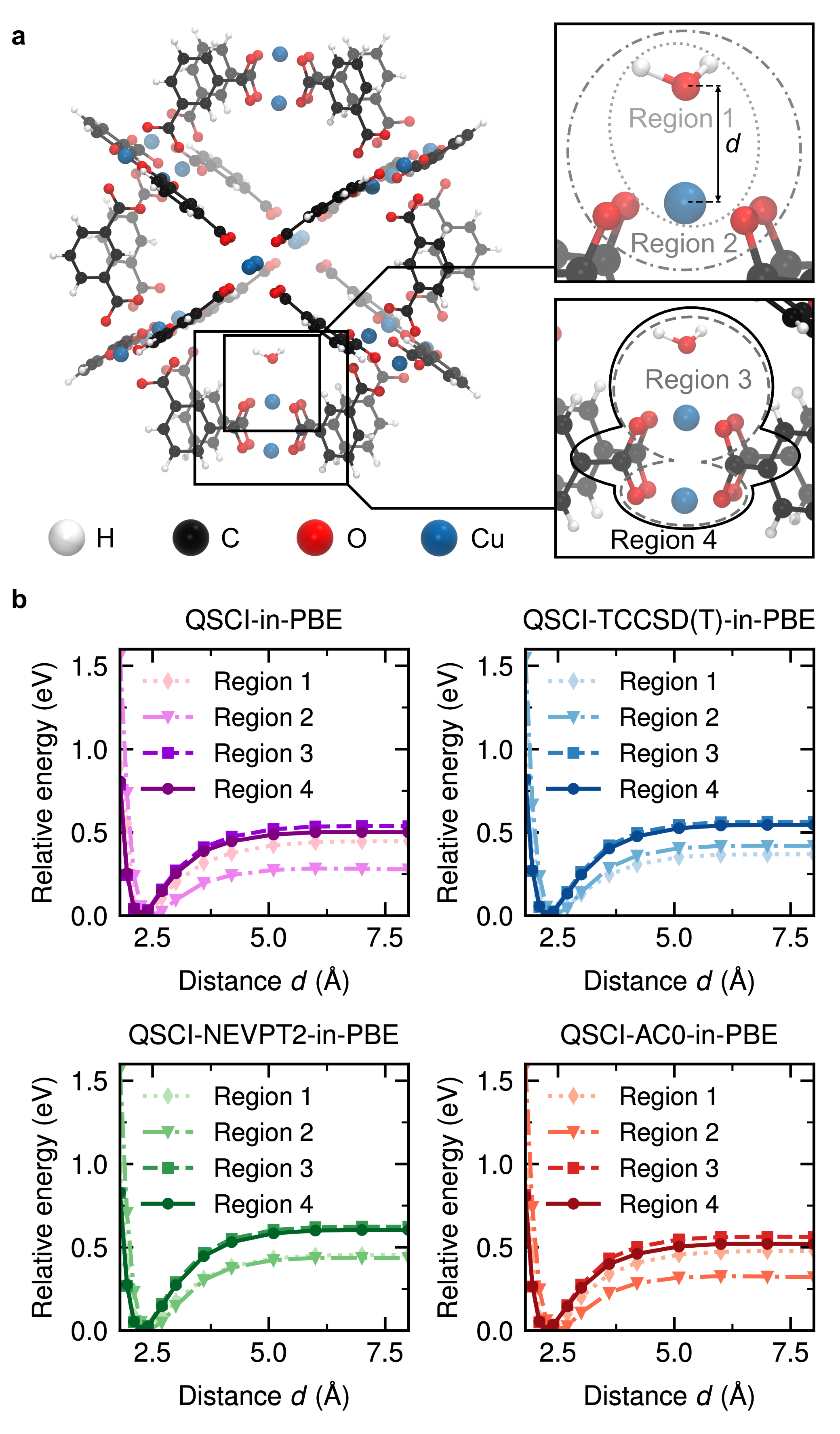}
    \centering
    \caption{\textbf{Results for water adsorption on an HKUST-1 MOF cluster model.} (a) Illustration of the HKUST-1 cluster model with a single water molecule adsorbed on a Cu site. Four different definitions for the active WF subsystem (Regions~1--4) were investigated. (b) Relative energy profiles along the reaction path computed with the SVP basis set. QSCI calculations used an active space of (2e,2o).}
    \label{fig:MOF_results}
\end{figure}

Next, we consider water adsorption on a cluster model of the HKUST-1 metal-organic framework (MOF).
MOFs are porous materials built from inorganic nodes connected by organic linker molecules. Their exceptionally high surface area, even compared to traditional porous materials such as zeolites, makes them highly attractive for applications in fuel storage, gas separation, and catalysis \cite{furukawaChemistryApplicationsMetalOrganic2013}.
HKUST-1, \ce{Cu3(BTC)2} (BTC = benzene-1,3,5-tricarboxylate) \cite{chuiChemicallyFunctionalizableNanoporous1999}, is a prototypical MOF for such applications and has therefore been extensively studied.
The adsorption of water on the Cu sites in HKUST-1 is of particular interest since it is closely linked to the stability of the framework and its performance in adsorption-based applications, such as gas separation and gas storage \cite{schlesingerEvaluationSyntheticMethods2010, grajciarWaterAdsorptionCoordinatively2010, giovineSurprisingStabilityCu3btc22018, xueTheoreticalInsightsInitial2020}.
We demonstrate that our quantum--classical approach can treat this adsorption process with systematically improvable accuracy.

Fig.~\ref{fig:MOF_results}a shows the HKUST-1 cluster model used in our simulations with a water molecule adsorbed on a Cu site. This finite fragment is structurally equivalent to the metal-organic polyhedron synthesized by Eddaoudi et al. \cite{eddaoudiPorousMetalOrganicPolyhedra2001}. 
We assessed four definitions for the active WF subsystem. Region~1 comprises the electrons and orbitals localized on the water molecule and the Cu site. Region~2 extends the subsystem by adding the nearest coordinating O atoms. Region~3 further includes the Cu and O atoms beneath the adsorption site, and Region~4 also incorporates C atoms connecting the O atoms and the aromatic rings. The remaining part of the MOF cluster constitutes the DFT environment treated at the PBE level. 

Fig.~\ref{fig:MOF_results}b presents the adsorption energy curves obtained with Regions~1--4 as the active WF subsystem. While QSCI-in-PBE yields smooth energy curves, the dependence on the chosen subsystem shows no clear trend for Regions~1--3: extending from Region~1 to Region~2 leads to a large decrease in the adsorption energy, whereas further extension to Region~3 raises it again above the Region~1 curve. Post-processing with TCCSD(T) or NEVPT2 restores a clear, systematic trend: going from Region~1 to Region~2 now leads to a small change, whereas extending from Region~2 to Region~3 produces a pronounced shift. This is consistent with the different WF-subsystem sizes and the inclusion of the second Cu atom directly beneath the adsorption site in Region~3, which may contribute strongly to the adsorption energy. QSCI-AC0-in-PBE, on the other hand, follows the behavior of QSCI-in-PBE, suggesting that AC0 post-processing is less effective at restoring the missing correlation in this case. In all methods, extending from Region~3 to Region~4 leads to a minor change, indicating convergence with respect to the WF-subsystem size.

For Region~4, the resulting adsorption energies and equilibrium distances fall in a reasonable range. We define the adsorption energy as the difference between the energies at the minimum and at \SI{8}{\angstrom}, which is approximately the pore center of the cluster model. The adsorption energies from the post-processed QSCI-in-PBE calculations are \SI{-0.55}{eV}, \SI{-0.62}{eV}, and \SI{-0.52}{eV} for TCC, NEVPT2, and AC0, respectively. These results are consistent with previously reported values for the periodic HKUST-1 MOF of \SI{-0.53}{eV} \cite{watanabeMolecularChemisorptionOpen2010} from DFT calculations with the PW91-GGA exchange-correlation functional and \SI{-0.57}{eV} \cite{grajciarWaterAdsorptionCoordinatively2010} from DFT/CC calculations \cite{bludskyInvestigationBenzenedimerPotential2008}. The equilibrium distance is \SI{2.25}{\angstrom} for all post-processed calculations, which is also in good agreement with the experimental value of \SI{2.17}{\angstrom} \cite{chuiChemicallyFunctionalizableNanoporous1999}.
Note that we chose a cluster model because our current implementation does not yet support periodic boundary conditions. A periodic formulation would avoid truncation artifacts and capture contributions from periodic images, which could further improve agreement with experiments.

\subsection{CO hopping on an IrPdPtRhRu HEA nanoparticle}

\begin{figure}
    \includegraphics[width=0.48\textwidth]{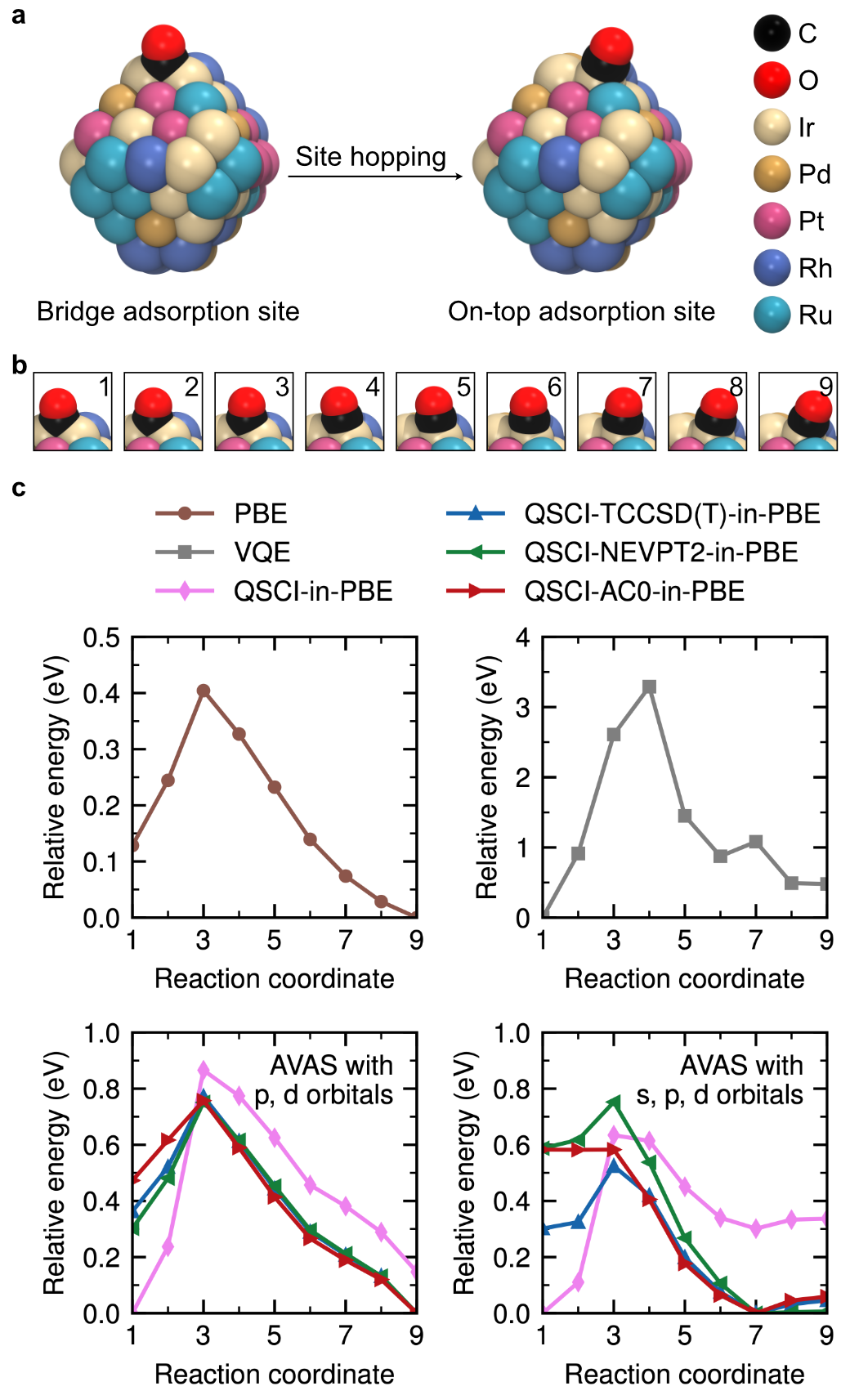}
    \centering
    \caption{\textbf{Results for CO hopping on an IrPdPtRhRu HEA nanoparticle.} (a) Illustration of CO hopping from a bridge to an on-top site on the IrPdPtRhRu HEA nanoparticle used in our simulations. (b) Geometries along the reaction path obtained from NEB-DFT calculations. (c) Relative energy profiles along the reaction path computed using the def2-SVP basis set with ECPs for the nanoparticle and the SVP basis set for the CO molecule. QSCI calculations used an active space of (4e,4o).}
    \label{fig:HEA_results}
\end{figure}

For the final application, we study CO hopping from a bridge site to an on-top site on an IrPdPtRhRu HEA nanoparticle.
HEAs are multicomponent alloys with five or more principal elements. Their tunable electronic structure can lead to exceptional physical and chemical properties, which have attracted wide attention in catalysis and other areas of chemistry and materials science \cite{yehNanostructuredHighEntropyAlloys2004,cantorMicrostructuralDevelopmentEquiatomic2004,georgeHighentropyAlloys2019,wuPlatinumGroupMetalHighEntropyAlloyNanoparticles2020,duAlloyElectrocatalysts2023,ouyangRiseHighentropyBattery2024,shiotaLoweringExponentialWall2025, caiSurfaceengineeredNanostructuredHighentropy2025,doOptimizingAdsorptionConfigurations2025}.
To assess the catalytic performance of HEAs, the adsorption properties of small molecules such as CO on HEA surfaces are often used as practical descriptors \cite{salinas-quezadaUnderstandingCOOxidation2024, shiotaLoweringExponentialWall2025}. We show that our quantum--classical method can be used to obtain systematically improvable energy profiles with reaction barriers in a physically reasonable range.

Fig.~\ref{fig:HEA_results}a illustrates CO hopping on the 79-atom IrPdPtRhRu HEA nanoparticle simulated in this work, and Fig.~\ref{fig:HEA_results}b shows the positions of the CO molecule along the reaction coordinate. We consider a truncated octahedron in a face-centered-cubic (fcc) arrangement for the nanoparticle. The chosen shape and size are realistic, as HEA nanoparticles with an average diameter of approximately \SI{1.32}{nm} have been synthesized \cite{wuPlatinumGroupMetalHighEntropyAlloyNanoparticles2020,minamiharaContinuousFlowReactorSynthesis2022}. The active WF subsystem includes the electrons and orbitals on the C, O, and Ir atoms involved in the hopping, while the remainder forms the DFT environment described at the PBE level. 

The relative energy curves are shown in Fig.~\ref{fig:HEA_results}c. PBE predicts the bridge site to be energetically higher than the on-top site. The maximum along the reaction coordinate occurs at the third geometry with a relative energy of \SI{0.4}{eV}. This is consistent with prior studies of CO adsorption on Ir(111), which report the on-top site as the most stable adsorption site and diffusion barriers of $\sim$\SI{0.4}{eV} \cite{krekelbergAtomicMolecularAdsorption2004, noeiMonitoringInteractionCO2018, liCOAdsorptionDisproportionation2022}.

Before turning to our QSCI-based results, we briefly discuss a VQE calculation performed on a simulator with an active space of (4e,4o). While VQE is not the focus of this work, it serves as an instructive example for demonstrating how challenging this barrier is to describe using quantum algorithms without DFT embedding and classical post-processing. The relative energy profile obtained with VQE gives the wrong relative stability of the adsorption sites, placing the bridge site below the on-top site. Furthermore, while the position of the maximum is shifted only slightly to the fourth geometry, its value is almost an order of magnitude larger than PBE, reaching \SI{3.3}{eV}.

We now show how DFT embedding and post-processing recover physically reasonable relative energy profiles. When selecting only $p$ and $d$ orbitals for the active WF subsystem via AVAS, QSCI-in-PBE reduces the maximum along the reaction coordinate to \SI{0.9}{eV} but predicts the wrong relative stability between the bridge and on-top sites. After post-processing with TCCSD(T), NEVPT2, or AC0, the bridge site correctly lies above the on-top site, and the value of the maximum is \SI{0.8}{eV}. 
When $s$ orbitals are included in the selection for the active WF subsystem, QSCI-in-PBE reduces the maximum further to \SI{0.6}{eV} but raises the relative energy of the on-top site. 
While all three post-processing methods recover the correct relative stability between the bridge and on-top sites, only TCCSD(T) and NEVPT2 show a clear maximum along the reaction coordinate, amounting to \SI{0.5}{eV} and \SI{0.8}{eV}, respectively.

Note that localizing orbitals in nanoparticles is itself a challenge due to the presence of many near-degenerate, entangled states. Several approaches have been proposed to address this and improve robustness, including maximally localized Wannier functions ~\cite{marzariMaximallyLocalizedGeneralized1997} with disentanglement ~\cite{souzaMaximallyLocalizedWannier2001} and density-matrix-based approaches such as selected columns of the density matrix ~\cite{damleCompressedRepresentationKohn2015} and its variants ~\cite{damleSCDMkLocalizedOrbitals2017, fuemmelerSelectedColumnsDensity2023}.

\section{Conclusions}
We introduced a quantum--classical framework that enables quantitative large-scale electronic-structure calculations where quantum computing is performed on the challenging, strongly correlated part of the system. In this framework, the sampling-based quantum algorithm QSCI connects the quantum treatment of the strongly correlated active space with the subsequent classical correlation treatments. By integrating projection-based WF-in-DFT embedding, our workflow addresses the scalability bottleneck of the classical post-processing step required to recover dynamical correlation beyond the active space treated by a quantum algorithm.

We performed numerical validations on O--H bond dissociation in ethanol and C$\equiv$N bond stretching in propionitrile, and applied our framework to the Menshutkin $\mathrm{S_N2}$ reaction inside a CNT, water adsorption on an HKUST-1 MOF cluster, and CO hopping on an IrPdPtRhRu HEA nanoparticle. Those calculations used a subset of qubits of a 144-qubit superconducting quantum computer at the University of Osaka.

Our results demonstrate that embedding QSCI in a DFT environment improved agreement with reference energies from high-level methods for the small benchmark molecules compared with the vanilla QSCI. For the large-scale systems, QSCI-in-DFT yielded energy profiles that captured the main qualitative features. To recover dynamical correlation beyond the active space, we evaluated TCC, NEVPT2, and AC0 as classical post-processing methods. All three methods achieved comparable accuracy for the small benchmark molecules. For the large-scale systems, TCC yielded the most accurate and robust results, achieving $\sim$\SI{1}{kcal/mol} agreement with the classical reference for the Menshutkin $\mathrm{S_N2}$ reaction inside a CNT. While NEVPT2 and AC0 also generally improved the results, their performance was more system dependent.

At the same time, our work shows that the reliability of the approach depends on the definition of the WF subsystem and that orbital localization can become challenging in electronically complex systems. Addressing these issues will be important for improving the robustness and predictive accuracy of the framework and for further advancing practical applications of quantum computing to problems in catalysis, materials science, and other areas where environmental effects are often essential.

\section{Acknowledgments}
The authors are grateful to Toru Shiozaki at QSimulate for adding the
QSimulate-QM functionality used in this work to construct
the DFT-embedded Hamiltonians. The authors also thank Takafumi Miyanaga and Kosuke Miyaji for their continued support in preparing the execution environment and enabling cloud-based access to the quantum computer, and Takuma Murokoshi for technical support.
This work was supported by MEXT Quantum Leap Flagship Program (MEXTQLEAP) Grant No. JPMXS0120319794 and the JST COI-NEXT Program Grant No. JPMJPF2014, and the JST ASPIRE Program Grant No. JPMJAP2319. 
This research was partially supported by the JSPS Grants-in-Aid for Scientific Research (KAKENHI) Grant No. JP23H03819.
\newpage

\bibliography{References}
\end{document}